\documentclass[12pt]{article}

\usepackage{graphicx}
\usepackage{amsmath}
\usepackage{amssymb}
\usepackage{wrapfig}
\usepackage{indentfirst}
\usepackage{color}
\usepackage{subfigure}
\usepackage{hyperref}

\usepackage{cite}

\begin{document}

\vskip 0.5cm \centerline{\bf\Large Investigating exclusive $\rho^0$ photoproduction } 
\centerline{\bf\Large within the Regge phenomenology approach}
\vskip 0.3cm
\centerline{L\'aszl\'o~Jenkovszky $^{a\star}$,  Érison S. Rocha$^{b\clubsuit}$ and Magno V. T. Machado$^{b\spadesuit}$}

\vskip 1cm

\centerline{$^a$ \sl Bogolyubov ITP,
National Academy of Sciences of Ukraine, Kiev
03143 Ukraine}
\centerline{$^b$ \sl HEP Phenomenology Group, CEP 91501-970, Porto Alegre, RS, Brazil}
\vskip
0.1cm

\begin{abstract}\noindent
The elastic differential and integrated total cross section for the exclusive $\rho^0$ photoproduction in electron-proton ($ep$) collisions are evaluated taking into account nonperturbative Pomeron exchange approach. By using three different models based on Regge phenomenology the results are compared to recent measurements by H1 Collaboration in $ep$ collisions and by the CMS collaboration from ultraperipheral proton-lead collisions. The analysis is expanded by calculating the coherent nuclear cross section, $\sigma (\gamma A\rightarrow \rho^0 A)$, which is applied to $\rho^0$ production in ultraperipheral lead-lead and xenon-xenon collisions. The predictions are compared to the measurements performed by ALICE Collaboration. Aspects of the theoretical uncertainties and limitations of the formalism are scrutinized.
\end{abstract}

\vskip 0.1cm

$
\begin{array}{ll}
^{\star}\mbox{{\it e-mail address:}} &
   \mbox{jenk@bitp.kiev.ua} \\
 ^{\clubsuit}\mbox{{\it e-mail address:}} &
\mbox{erison.rocha@ufrgs.br}  \\
 ^{\spadesuit}\mbox{{\it e-mail address:}} &
\mbox{magnus@if.ufrgs.br}\\  

\end{array}
$


\section{Introduction}\label{Int}

The exclusive $\rho^0$ photoproduction has been investigated recently both experimentally and theoretically \cite{H1:2020lzc,CMS:2019awk,ALICE:2020ugp,ALICE:2021jnv}. Light vector mesons have not a hardness scale associated to the process in the photoproduction limit. Therefore, they can be used to test the nonperturbative (soft) regime of the strong interactions. In the context of parton (gluon) saturation approaches the transition between the region described by perturbative Quantum Chromodynamics (pQCD) and the soft regime is interpreted in terms of the saturation momentum scale \cite{hdqcd,Morreale:2021pnn}, $Q_s(x)$ ($x$ is the Bjorken invariant). In the color dipole picture \cite{nik,Nemchik:1996pp} the saturation scale characterizes the boundary on the maximum phase-space gluon density to be reached in the wavefunction of the hadron. In principle, the light meson photoproduction dynamics at high energies would be driven by $Q_s$ as it reaches values $\lesssim 1$ GeV in this regime. The perturbative description is even better for the scattering on nuclei targets as the nuclear saturation scale, $Q_{s,A}^2\propto A^{1/3}Q_s^2$, is enhanced. Nevertheless, it is already known that such an approach is unable to describe correctly the total photoproduction cross section and $\rho^0$ production at $Q^2=0$ GeV$^2$ \cite{Goncalves:2020cir,Forshaw:1999uf}. In particular, nonperturbative corrections to the description are needed and usually they are introduced in the photon wavefunction for color dipoles of large transverse size.  

On the other hand, Regge phenomenology is historically appropriated formalism to describe soft and diffractive processes, Deeply Virtual Compton Scattering (DVCS) as well as light meson photoproduction. The production amplitude is written in a Regge-factorized structure with the corresponding coupling of particles to the Pomeron. The introduction of a hard scale dependence suitable for vector meson electroproduction can be carefully constructed based on geometric arguments. The Reggeometric Pomeron model \cite{Fazio:2013uwa,Fazio:2013hza} is one example of such a class of phenomenological formalism. The approach does a good job in describing the vector meson photo and electroproduction at the DESY-HERA energy regime considering a nucleon target. The possibility for testing these models in the coherent vector meson production in ultraperipheral heavy ion collisions (UPCs) is now a reality. The basic idea is that the production cross section in $AA$ collisions can be factorized in terms of the equivalent flux of photons of the colliding nucleus and the photon-target production cross section \cite{Klein:2020fmr}.

In this work the $\rho^0$ photoproduction in $ep$ and $AA$ collisions will be examined closely. Three different models based on Regge phenomenology \cite{Jenkovszky:2018itd} are investigated: the Reggeometric Pomeron (RP) model \cite{Fazio:2013uwa,Fazio:2013hza}, the Soft Dipole Pomeron (SDP) \cite{Jenkovszky:1996hk,Fiore:1998jx,Martynov:2002ez} and a factorized Regge model with nonlinear Pomeron trajectory (NL) \cite{Fazio:2011ex}. In $ep$ collisions the elastic differential cross section and the integrated total cross section will be analysed. The results are compared to recent measurements performed by HERA-H1 \cite{H1:2020lzc} and CMS collaborations \cite{CMS:2019awk}. Making used of Vector Dominance Model (VDM) and Glauber multiple scattering formalism the coherent cross section is computed. The results are compared to the measurements for $\rho^0$ production in PbPb and XeXe UPCs done by ALICE Collaboration at the Large Hadron Collider (LHC) \cite{ALICE:2020ugp,ALICE:2021jnv}. Namely, the rapidity distribution has been determined in PbPb UPCs at $\sqrt{s_{\mathrm{NN}}}=5.02$ TeV and XeXe UPCs at $\sqrt{s_{\mathrm{NN}}}=5.44$ TeV. The present investigation extends our earlier works~\cite{Fiore:2014oha,Fiore:2014lxa,Fiore:2015yya,Jenkovszky:2021sis,Jenkovszky:2022wcw} where the rapidity distribution and $|t|$-dependence of quarkonium production in proton-proton and UPCs in $AA$ collisions have been addressed. The investigation is also relevant for the physics of the Electron Ion Collider (EIC) \cite{Accardi:2012qut}, Electron-ion collider in China (EicC) \cite{Anderle:2021wcy} and the Large Hadron Electron Collider (LHeC) \cite{Andre:2022xeh}. The topic is matter of intense theoretical studies in recent years \cite{Henkels:2022bne,Guzey:2020pkq,Ma:2019mwr,Guzey:2018bay,Goncalves:2018blz,Xie:2022sjm,Xing:2020hwh,Demirci:2022wuy,Cisek:2022yjj,Mantysaari:2022bsp,Cepila:2018zky,Bendova:2018bbb,Khoze:2019xke}. 

This paper is organized as follows. In Sec. \ref{sec:theory} we briefly review the exclusive $\rho^0$ meson production, $\gamma+p\rightarrow  \rho^0+p$, in the context of the Reggeometric, Soft Dipole and nonlinear trajectory Pomeron models. Afterwards, using VDM approach and Glauber formalism for nuclear effects, the expression for the coherent nuclear cross section is revised. In section \ref{sec:results} the theoretical calculation is compared to available experimental measurements in $ep$ and $AA$ UPCs collisions at the LHC. A discussion on the theoretical uncertainties is done. In the last section the main results are summarized.

\begin{figure}[t]
\centering
\includegraphics[width=.8\textwidth]{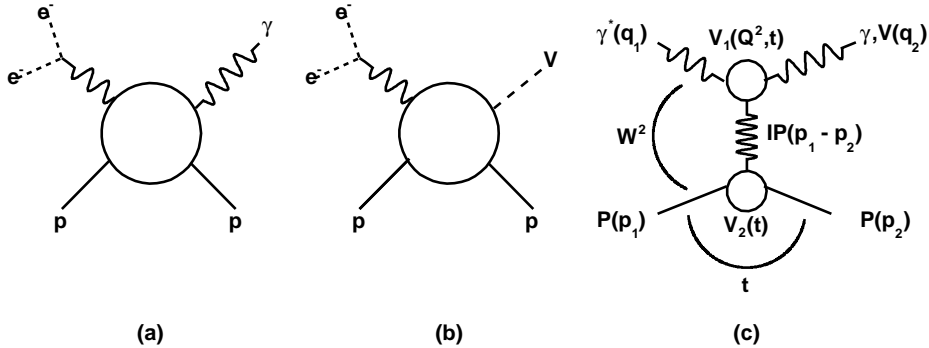}
\caption{Diagrams of DVCS (a) and vector meson production  (b) in lepton-nucleon scattering; (c) DVCS/meson production amplitude in a Regge-factorized structure including representation for particle vertices, $V_{1}(Q^2,t)$ and $V_2(t)$.}
\label{fig:diagrams}
\end{figure}

\section{Theoretical framework}
\label{sec:theory}
Exclusive $\rho^0$ photoproduction process, $\gamma +p\rightarrow \rho^0+p$, will be described by using three models based on Regge phenomenology (see Fig. \ref{fig:diagrams}-b). The first one is the Reggeometric Pomeron (RP) model, the second is the Soft Dipole Pomeron (SDP) model and the third one is a factorized Regge model with nonlinear Pomeron trajectory (NL). All of them  are also able to describe electroproduction data. In general case, the hardness parameter is given by $\widetilde Q^2=Q^2+M_V^2$, where $Q^2$ is the photon virtuality and $M_V$ is the vector meson mass.   

The elastic differential cross section, $d\sigma_{el}/dt$, and the total cross section, $\sigma (\gamma^*+p\rightarrow \rho^0+p)$, associated to the single-component RP model in a given scale $\widetilde Q^2$ are given by \cite{Fazio:2013uwa,Fazio:2013hza}:
\begin{eqnarray}
\frac{d\sigma_{el}^{\mathrm{RP}}}{dt}& =& \frac{A_0^2}{\left(1+\frac{\widetilde{Q^2}}{{Q_0^2}}\right)^{2n}}\left(\frac{W_{\gamma p}^2}{W_0^2}\right)^{2(\alpha(t)-1)}\exp \left[B_0(\widetilde{Q^2})\,t\right], \quad B_0(\widetilde{Q^2})= 4\left(\frac{a}{\widetilde{Q^2}}+\frac{b}{2m_N^2}\right), \\
\sigma_{\gamma p \to V p}^{\mathrm{RP}}  & = &\frac{A_0^2}{\left(1+\frac{\widetilde Q^2}{\widetilde Q^2_0}\right)^{2n}}\frac{\left(W_{\gamma p}/W_0\right)^{4(\alpha_0-1)}}{B\left(W_{\gamma p},\widetilde Q^2 \right)}, \quad B \left(W_{\gamma p},\widetilde Q^2 \right)  =  B_0(\widetilde{Q^2})+ 4\alpha'\ln \left(\frac{W_{\gamma p}}{W_0}\right),
\label{reggeometric-exp}
\end{eqnarray}
where the quantity $B_0(\widetilde{Q^2})$ reflects the geometrical nature of the model and $\alpha (t)$ denotes the effective Pomeron trajectory. The first and second term in $B_0(\widetilde{Q^2})$ correspond to the effective sizes of upper and lower vertices in Fig. \ref{fig:diagrams}-c, respectively. In the formulas above, $W_0=1$ GeV.
 
\begin{table}[t]
  \centering
   \footnotesize
   \caption{Values of the parameters for the Reggeometric Pomeron model \cite{Fazio:2013hza} fitted to data on $\rho^0$ production measured by DESY-HERA \cite{H1:1996prv,ZEUS:1997rof,ZEUS:1995bfs}. The recent high precision data of Ref. \cite{H1:2020lzc} are not included.}
   \label{tab:1}
  \begin{tabular}{|c|c|c|c|c|c|c|}
     \hline
                 $A_0$ $\left[\frac{\sqrt\text{nb}}{\text{GeV}}\right]$
                 &$\widetilde{Q^2_0}$ $\left[\text{GeV}^2\right]$&   $n$
                 &$\alpha_{0}$& $\alpha'$  $\left[\text{GeV}^{-2}\right]$
                 &$a$&$b$ \\ \hline 
         344 $\pm$ 376 & 0.29 $\pm$ 0.14      &1.24 $\pm$ 0.07 &1.16 $\pm$ 0.14 & 0.21 $\pm$ 0.53& 0.60 $\pm$ 0.33 & 0.9 $\pm$ 4.3  \\ 
 \hline
        \end{tabular}
 \end{table}

In the photoproduction limit $\tilde{Q}^2=M_V^2$. The original parameters of the model are presented in Table \ref{tab:1}, which were determined in Ref. \cite{Fazio:2013hza} by using previous DESY-HERA measurements \cite{H1:1996prv,ZEUS:1997rof,ZEUS:1995bfs}. It should be noticed that the quality of fit $\chi^2/\mathrm{dof}=2.74$ for the $\rho^0$ meson is not optimal and authors called attention that a single $e^{Bt}$ parametrization is not sufficient to reproduce the $d\sigma/dt$ above $|t| > 0.5$ GeV$^2$ in both electroproduction and photoproduction regions. It is symptomatic the large uncertainty on the parameters around the central value and the significantly large value of the effective Pomeron intercept, $\alpha_0=1.16$.

\begin{table}[t]
  \centering
   \footnotesize
   \caption{Values of the parameters for the Soft Dipole Pomeron model \cite{Jenkovszky:1996hk} fitted to data on $\rho^0$ production at HERA \cite{H1:1996prv,ZEUS:1997rof,ZEUS:1995bfs}. }
   \label{tab:2}
  \begin{tabular}{|c|c|c|c|c|}
     \hline
                 $A_1$ $\left[\frac{\sqrt{\mu\text{b}}}{\text{GeV}}\right]$
                 &$A_2$ $\left[\frac{\sqrt{\mu\text{b}}}{\text{GeV}}\right]$&   $b_1$ [GeV$^{-2}$]
                 &$b_2$  [GeV$^{-2}$]& $s_0$ $\left[\text{GeV}\right]$ \\ \hline 
         -13.944 & 8.3626     & 3.89  & 1.83 & 1.00 \\ 
 \hline
        \end{tabular}
 \end{table}
 
Now, the differential cross section and total cross section for the SDP model is presented \cite{Jenkovszky:1996hk,Fiore:1998jx,Martynov:2002ez}. These observables are expressed as \cite{Jenkovszky:1996hk}:
\begin{eqnarray}
\frac{d\sigma_{el}^{\mathrm{SDP}}}{dt}&=&\left(\frac{s}{s_0}\right)^{2\alpha(t)-2}\left\{\left[G_1(t)+G_2(t)\ln{\left(\frac{s}{s_0}\right)}\right]^2+
\frac{\pi^2}{4} G_2^2(t)\right\}, \\
\sigma_{\gamma p \to V p}^{\mathrm{SDP}}  & = & 4\pi \left\{ \frac{A_1^2}{2B_1} + 2\frac{A_1A_2}{B_1+B_2}\ln\left(\frac{W_{\gamma p}^2}{W_0^2} \right) + \frac{A_2^2}{2B_2}\left[\ln^2\left(\frac{W_{\gamma p}^2}{W_0^2} \right)  + \frac{\pi^2}{4} \right] \right\} ,
\label{sigmasdp}
\end{eqnarray}
where the residua of the single, $G_1$, and double pole, $G_2$, are parametrized as follows,
\begin{eqnarray}
G_1(t)=A_1e^{b_1t}g_1(t), \quad G_2(t)=A_2e^{b_2t}g_2(t).
\label{residuos}
\end{eqnarray}
 In the phenomenology, $G_1(t)$ factorizes into a usual $pI\!\!Pp$-vertex, $G_1\propto e^{b_1t}$, and a $\gamma I\!\!PV$-vertex, $g_1(t)$, which in general is parametrized by a polynomial on $t$. The residue of the double pole $G_2$ can be cast from $G_1$, but such a correlation can be relaxed. For instance, in Ref. \cite{Fiore:1998jx} one uses $b_1=b_2\sim 2.25$ GeV$^{-2}$, $g_1 = (1+h_1t)$ and $g_2 \approx (1+h_2t)$. In order to obtain Eq. (\ref{sigmasdp}) the simplest option has been considered following Ref. \cite{Jenkovszky:1996hk}, i.e. $g_1(t)=g_2(t)=1$. This is enough for the present study. Here, a linear trajectory for the Pomeron, $\alpha(t)=1 + \alpha^{\prime} t$ is considered, with $\alpha^{\prime}=0.25$ GeV$^{-2}$. Moreover, in Eq. (\ref{sigmasdp}) the following notation is employed, $B_i = b_i+\alpha^{\prime}\ln (W_{\gamma p}^2/W_0^2)$, with $i=1,2$ and $W_0 = 1$ GeV.
 
 The parameters in Ref. \cite{Jenkovszky:1996hk} are fitted including the single Pomeron exchange, the $f_2$ contribution and the $I\!\!P\!-\!f_2$ interference. Here, only the $I\!\!P$-exchange is taken into account and the parameters $A_{1,2}$ were tuned to describe the old DESY-HERA data \cite{H1:1996prv,ZEUS:1997rof,ZEUS:1995bfs} whereas $b_{1,2}$ are fixed as their original values ($b_1 = 3.89$ GeV$^{-2}$ and $b_2 = 1.83$ GeV$^{-2}$ ). The values of the parameters for $\rho^0$ are presented in Table \ref{tab:2}.

Finally, a Regge model taking into account nonlinear Pomeron trajectory (NL) \cite{Fazio:2011ex} is investigated. It is a factorized Regge pole model where the dependence on the mass and virtuality of the
external particles enters via the relevant residue functions. On the other hand, the Pomeron trajectory is universal and scale-independent with the form $\alpha(t)  = \alpha_0-\alpha_1\ln(1-\alpha_2t)$. Here, the $\alpha_i$ are the $\alpha (t)$-trajectory parameters. It produces an approximately linear behaviour at small $|t|$, with $\alpha^{\prime} = \alpha_1\alpha_2$ being the forward slope. Moreover, at large $|t|$ the cross section presents scaling behaviour driven by the quark counting rule \cite{Fazio:2011ex}. The invariant scattering amplitude and the differential cross section are written as:
\begin{eqnarray}
A(s,t,\widetilde{Q}^2)_{\gamma^* p \rightarrow V p}^{\mathrm{NL}}=-A_0\,4\pi\,V_1(t,\widetilde{Q}^2)\,V_2(t)\,\left(\frac{-is}{s_0}\right)^{\alpha(t)-1},\quad \frac{d\sigma_{el}^{\mathrm{NL}}}{dt} = \frac{1}{16\pi} \left|A(s,t,\widetilde{Q}^2)  \right|^2,
\label{nonlinamp}
\end{eqnarray}
where $A_0$ is the normalization factor, $V_1(t,\widetilde Q^2)= \exp[b_2\beta(z)]$ is the $\gamma I\!\!PV$-vertex and $V_2(t)=$ exp$[b_1\alpha(t)]$ is the $pI\!\!Pp$-vertex. The quantities $\beta(z)$ and $\alpha(t)$ are the exchanged Pomeron trajectory in the meson vertex and in the proton vertex, respectively. In Ref. \cite{Fazio:2011ex} the vertex $V_1$ has been modeled in such way that it contains $t$-dependence and also $Q^2$-dependence through the trajectory, $\beta(z) = \beta_0-\beta_1\ln(1-\beta_2z)$, where $z = t-\widetilde Q^2$. The constants $\beta_i$ are the $\beta(z)$-trajectory parameters. Accordingly, $\beta_1\beta_2=\beta'$ is the forward slope of this trajectory. In exclusive vector meson production, the hardness in this model is the same as for the Reggeometric Pomeron one, $\widetilde Q^2 = Q^2+M_V^2$.

The calculation of the integrated cross section with the nonlinear trajectory from the amplitude in Eq.~(\ref{nonlinamp}) produces the following expression in the limit $s \rightarrow \infty$ and negligible nucleon mass \cite{Fazio:2011ex}:
\begin{eqnarray}
\sigma(s,\tilde Q^2)^{\mathrm{NL}}_{\gamma^* p \rightarrow V p}=\int^{\infty}_0~d|t|~\frac{d\sigma(s,t,\tilde Q^2)}{dt}=\frac{K}{\mu-1}{~_2}F_1\left(\nu,\mu-1;\mu;1-\beta \right),
\label{eq:NL}
\end{eqnarray}
where  $_{~2}F_1\left(\nu,\mu-1;\mu;1-\beta \right)$ is the Gauss hypergeometric function. The remaining quantities appearing in the equation above are defined by: 
\begin{eqnarray}
K &=& \frac{\pi|A_0|^2}{\alpha_2}e^{2\alpha_0(b_1+b_2)}(s/s_0)^{2\alpha_0-2}, \quad \mu = 2\alpha_1\left[b_1+b_2+\ln(s/s_0) \right]\\
\beta &=& 1+\alpha_2 \tilde Q^2, \quad \nu = 2b_2\alpha_1.
\end{eqnarray}

The fixed parameters of the model are $\alpha_0=\beta_0=1.09$, $\alpha_1=\beta_1=2.00$, $s_0=1$ GeV$^2$, $\alpha_2=\beta_2= 0.125$ GeV$^{-2}$ and $b_1=2.00$, which are constrained by plausible assumptions \cite{Fazio:2011ex}. There are only two free parameters: the parameter $b_2$, entering the $\gamma I\!\!PV$-vertex and the normalization factor, $|A_0|^2$. In the the original paper \cite{Fazio:2011ex}, they were only fitted to the DESY-HERA electroproduction data. Here, we have tuned them to describe the $\rho^0$ photoproduction data measured in Refs. \cite{H1:1996prv,ZEUS:1997rof,ZEUS:1995bfs}. One obtains the values, $b_2 = 2.300$ and $|A_0|^2 = 0.885$ nb. It was checked that the new tuned values are consistent with the original ones, for instance $b_2=1.087$ and $|A_0|^2 = 0.839$ nb at $Q^3 = 3.3$ GeV$^{2}$. Of course, the factor two of enhancement on $b_2$ parameter will bring consequences for the $t$-slope in the photoproduction limit. 

We are also interested in the nuclear coherent production of $\rho^0$, which is relevant for the physics at the EIC, EicC and LHeC \cite{Accardi:2012qut,Anderle:2021wcy,Andre:2022xeh}. Moveover, the coherent production is the main input for vector meson production in ultraperipheral heavy ion collisions. We follow the Klein-Nystrand approach \cite{Klein:1999qj,Nystrand:1998hw} which is the basis for the STARLIGHT Monte Carlo generator for UPCs processes \cite{Klein:2016yzr}.  The nuclear effects for the process, $\gamma +A\rightarrow \rho^0+A$ are described here by vector dominance model (VDM) \cite{Bauer:1977iq} and the classical mechanics Glauber (CM) formula for multiple scattering of the $\rho^0$ in the
nuclear medium. The differential cross section at $t=0$ is obtained by using the optical theorem for scattering in a nucleus and VDM in the following way \cite{Klein:1999qj,Nystrand:1998hw,Klein:2016yzr}:
\begin{eqnarray}
\left. \frac{\mathrm{d}\sigma\left( \gamma + A \to \rho^0 + A \right)}{\mathrm{d}t}\right|_{t=0}
 =  \frac{\alpha_{em}}{4 f^2_{\rho}}\sigma_{tot}^2\left( \rho^0 A \right)= \frac{\alpha_{em}}{4 f^2_{\rho}}\left\{\int \mathrm{d}^2 \textbf{b} 
\left[ 1-\exp\left( -\sigma_{tot}\left( \rho^0 p \right) T_A\left(\textbf{b} \right) \right) \right]\right\}^2,
\label{glauberVMD}
\end{eqnarray}
where $T_A(b)$ is the nuclear thickness function and $f_{\rho}$ is the vector-meson coupling. The value $f^2_{\rho}/4\pi = 2.02$ is considered in calculations. The formalism above has been extended in the generalized vector dominance model \cite{Frankfurt:2003wv} (or Gribov-Glauber model), which takes into account coherence effects for the soft interactions for production of
$\rho$ and $\rho^{\prime}$-mesons. \textbf{Once $\sigma_{tot}(\rho^0p)$ is relatively large, the cross section $\sigma_{tot}(\rho^0A)\approx \pi R_A^2$ is approximately the  geometric cross section  which is almost energy independent.}

The input for the Glauber model calculation in Eq. (\ref{glauberVMD}) is the effective $\rho^0$–nucleon cross  for the process $\rho^0+p\rightarrow \rho^0+p $, which is given by:
\begin{eqnarray}
\sigma_{tot}\left( \rho^0 p \right)= \sqrt{\frac{4f^2_{\rho}}{ \alpha_{em}} \left.
\frac{\mathrm{d}\sigma\left( \gamma + p \to \rho^0 + p
  \right)}{\mathrm{d}t}\right|_{t=0}},
  \label{sigrhop}
\end{eqnarray}
where the predictions from the three phenomenological models will be introduced in Eq. (\ref{sigrhop}) .
The corresponding integrated cross section is expressed as:
\begin{eqnarray}
\sigma (\gamma + A \rightarrow \rho^0 + A) = 
\left. \frac{d\sigma (\gamma + A \rightarrow  \rho^0 + A)}{dt}\right|_{t=0}
\int\limits_{t_{min}}^{\infty} \mathrm{d}|t| \, \left| F_A\left(t\right) \right|^2, 
\label{eq:sigtotgammaA}
\end{eqnarray}
where $F_A$ is the nuclear form factor. An analytic form factor given by a hard sphere of radius, $R_A =r_0A^{1/3}$ fm, convoluted with a Yukawa potential with range $a$ \cite{Davies:1976zzb} will be considered. This is sufficient for practical utilization as shown in Ref. \cite{Klein:1999qj}. The analytical expression is the following,
\begin{eqnarray}
 F_A(|q|)  =  \frac{4\pi\rho_0}{A |q^3|} \left( \frac{1}{1+a^2q^2} \right) \left[ \sin{(|q| R_A)} - |q| R_A\cos{(|q| R_A)}  \right], 
\end{eqnarray}
where $A$ is the mass number, $q$ is the momentum transfer, $\rho_0 = 3/(4\pi r_0)$ fm$^{-3}$ and $a = 0.7$ fm.

Here, some discussion is needed on the underlying approximations present in Eq. (\ref{glauberVMD}). It is considered  the inelastic $\rho^0$–nucleus cross section instead of the total cross section which decreases the prediction for the forward cross section by a factor $\sim 2$. In the Klein-Nystrand approach (and STARLIGHT MC)  the total cross section of the $\rho^0A$ interaction is obtained from classical mechanics (MC) model. On the other hand, the quantum mechanics expression is given by the Gribov-Glauber (GG) formalism where the $\rho^0A$ cross section is given by:
\begin{eqnarray}
\sigma_{tot}^{\mathrm{GG}}(\rho^0A) = 2\int d^2\vec{b} \left[ 1-\exp \left( -\frac{1}{2}\sigma_{\rho^0N}T_A(\vec{b})\right) \right].
\end{eqnarray}
In the simplification of a sharp sphere nucleus with $\rho_0=0.17$ fm$^{-3}$ and radius $R_A$ one can obtain an estimate of the ratio between the GG and CM cross sections \cite{Frankfurt:2002sv}, 
\begin{eqnarray}
\frac{\sigma_{tot}^{\mathrm{GG}}(\rho^0A)}{\sigma_{tot}^{\mathrm{CM}}(\rho^0A) }\approx 2\left(1-\frac{3}{2\rho_0^2\sigma_{\rho^0N}^2R_A^2}  \right) .
\end{eqnarray}
The ratio is $\approx 1.67$ for lead ($R_A\simeq 7.1$ fm) and $1.55$ for xenon ($R_A\simeq 6.1$ fm) by using $\sigma_{\rho^0N}\approx 25$ mb. The classical probabilistic formula (CM) and the Glauber-Gribov (GG) approach give near values of the $\sigma_{tot}(\rho^0A)$ only when $\sigma_{tot}(\rho^0p) T_A(b) \ll 1$. A detailed discussion on the consequences of the approximations above presented can be found in Refs. \cite{Frankfurt:2002sv,Frankfurt:2002wc,Guzey:2013xba,Frankfurt:2015cwa,Guzey:2020pkq}.

In next section we investigate the $\rho^0$ photoproduction on nucleons and the nuclear coherent scattering by using as input the Regge models discussed before. The prediction will be compared to updated data sets. As a by product, the coherent production of $\rho^0$ at UPCs is analysed for both PbPb and XeXe collisions at the LHC.  

\begin{figure}[t]
\centering
 \includegraphics[width=.7\textwidth]{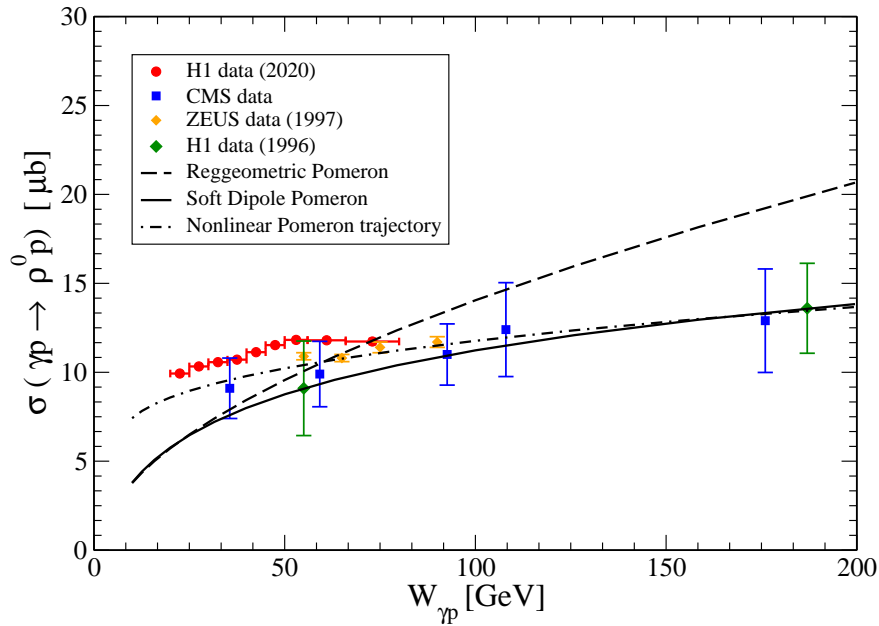}
 \caption{Cross section of exclusive $rho^0$ meson photoproduction as a function of centre of mass energy of the photon-nucleon system, $W_{\gamma p}$. Experimental measurements of H1, ZEUS \cite{H1:2020lzc,H1:1996prv,ZEUS:1997rof,ZEUS:1995bfs} and CMS \cite{CMS:2019awk} are presented. Reggeometric Pomeron prediction is represented by dashed curve, the Soft Dipole by solid curve and that one for Nonlinear Trajectory by the dot-dashed curve. }
 \label{fig:sigmaWH1-comp}
\end{figure}

\section{Results and discussions}
\label{sec:results}

Let us start with the numerical results and their comparison to recent experimental measurements for $\rho^0$ photoproduction by H1-HERA \cite{H1:2020lzc} and CMS Collaboration \cite{CMS:2019awk} at the LHC. In the last case, the the points has been extracted from ultraperipheral pPb coherent collisions at 5.02 TeV.

First, in Fig. \ref{fig:sigmaWH1-comp} the cross section of $\rho$ production in $\gamma p$ scattering is shown as a function of the photon-nucleon centre of mass energy. Previous measurements of the photoproduction cross section determined by H1 and ZEUS are also presented. The Reggeometric Pomeron model is labeled by the dashed-line, the Soft Dipole model by solid line and Nonlinear Pomeron trajectory by dot-dashed line. It should be stressed that the models parameters have been determined by fitting the old H1 and ZEUS data \cite{H1:1996prv,ZEUS:1997rof,ZEUS:1995bfs}. Here, a parameter free calculation is performed. The Reggeometric Pomeron model has the steeper energy dependence and the others presenting a milder growth. This is traced back to the effective Pomeron intercept in the former, $\alpha_0=1.16$ whereas in the Soft Dipole model the energy growth is proportional to $\ln^2 (W_{\gamma p}^2/W_0^2)$. In the nonlinear trajectory model a lower Pomeron intercept is considered, namely $\alpha_0=1.06$ compared to the Reggeometric one. As expected, once the parameters have been tuned to old HERA data, the previous measurements by H1 and ZEUS are better described. Given the relatively large error bars in CMS data, most of the models describe those measurements except the Reggeometric one at higher energies. It is worth mentioning that the $f_2$ contribution and its interference with the Pomeron was not considered here. However, these contributions can be important at the lower energy data points measured recently by H1 \cite{H1:2020lzc} . 
\begin{figure}[t]
\centering
 \includegraphics[width=.7\textwidth]{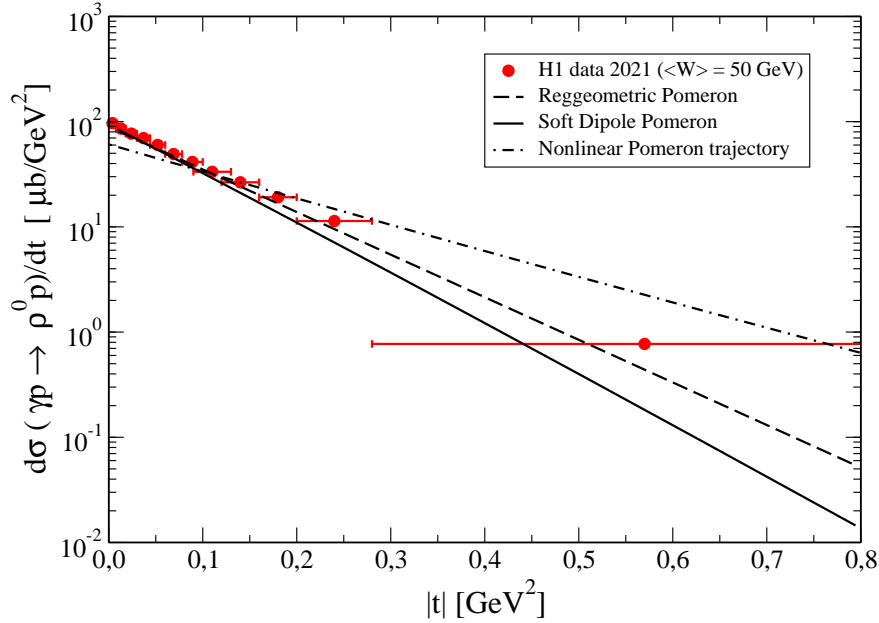}
 \caption{Differential cross section, $d\sigma /dt$, for exclusive $\rho^0$ photoproduction as a function of $|t|$ at fixed average energy $\langle W_{\gamma p}\rangle$ = 50 GeV measured by H1 Collaboration \cite{H1:2020lzc}. Same notation as previous figure.}
  \label{fig:dsdtxsectionHERA}
\end{figure}

\begin{figure}[t]
\centering
 \includegraphics[width=.9\textwidth]{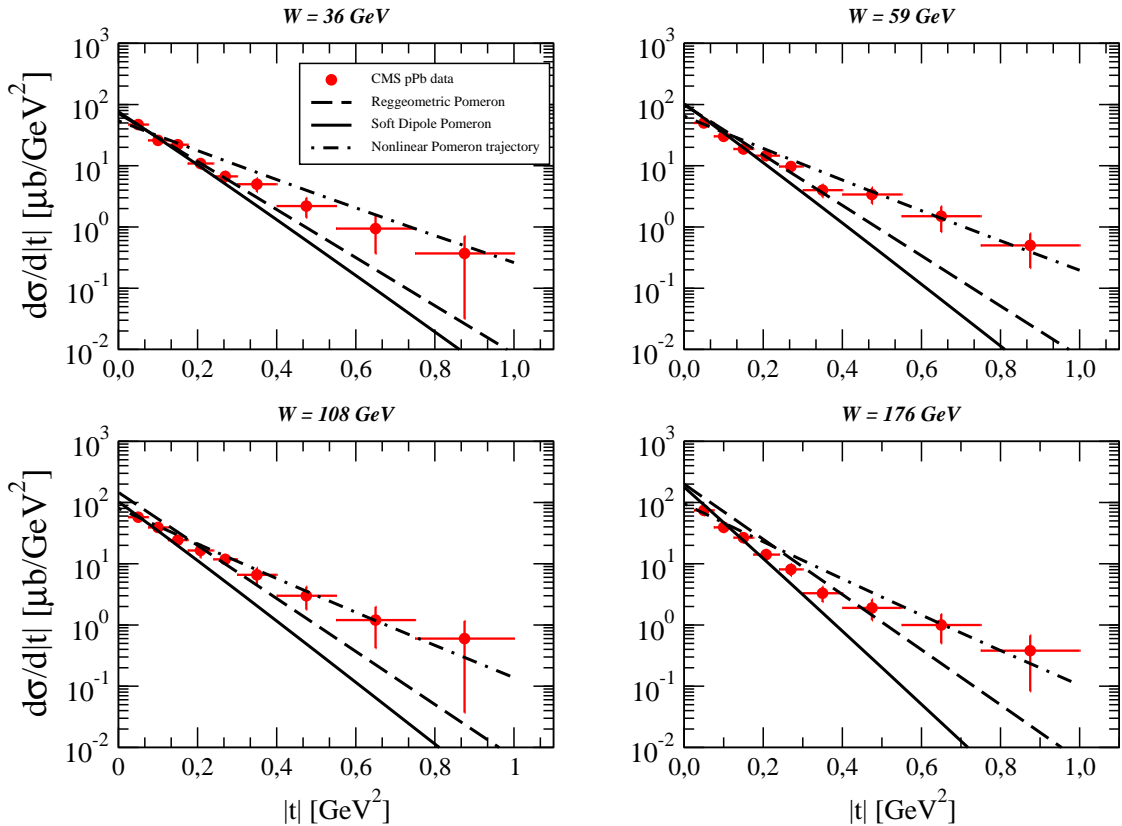}
 \caption{Differential cross section, $d\sigma /dt$, for exclusive $\rho^0$ photoproduction as a function of $|t|$ at fixed average energies of $\langle W_{\gamma p}\rangle$ = 35.6, 59.2, 108.6 and 176.0 GeV measured by CMS Collaboration \cite{CMS:2019awk}. Same notation as previous figure.}
  \label{fig:dsdtxsectionCMS}
\end{figure}

Secondly, the $|t|$-distribution is investigated and presented in Figs. \ref{fig:dsdtxsectionHERA} and \ref{fig:dsdtxsectionCMS}. The differential cross section is compared to the H1 data \cite{H1:2020lzc} in Fig. \ref{fig:dsdtxsectionHERA} at the average energy of $\langle W_{\gamma p} \rangle = 50$ GeV as a representative sample. As expected, both the Reggeometric and Soft Dipole models are able to describe the overall normalization and the $|t|$ slope at very small-$t$. At larger $t$ the description loses consistency and underestimate the central points at larger $t$. On the other hand, the present version of the nonlinear trajectory model underestimates the small-$t$ and overestimates the large $t$ data points. However, this can be just a consequence of the parameters tuning as discussed in previous section. It is urged a new high quality fits which include these new data for $\rho^0$ photoproduction in the analysis.

Third, in Fig. \ref{fig:dsdtxsectionCMS} the models are compared to the CMS Collaboration data \cite{CMS:2019awk}. The differential cross section has been measured in the interval $0.025<|t| < 1$ GeV$^2$ and for average values of energy, $\langle W_{\gamma p} \rangle  = 35.6,\,59.2,\,92.6, \,108.0$ and $176.0$ GeV, respectively. Concerning the Reggeometric and Soft Dipole models, both underestimate large $t$ data following the observed behavior against H1 data. Now, the NL model does a better job compared to the H1 data description. The CMS measurements cover a wider interval of $t$ mostly including larger values compared to the H1 ones. It has been reported in Ref. \cite{CMS:2019awk} that the STARLIGHT Monte Carlo prediction is widely
higher than the CMS data in the high-$t$ region and the deviations consistently increase with energy. 

At this point, the behavior can be traced back to the average slope coming out from each model. As an exercise, let us consider $\langle W_{\gamma p} \rangle =$ 92.6 GeV and compute the predicted $t$-slope by using the original parameters of the models. In the RP model, the slope is given by Eq. (\ref{reggeometric-exp}) and $B_{\mathrm{RP}} \left(\langle W_{\gamma p}\rangle, \widetilde{Q}=m_{\rho} \right)  \approx 9.83 $ GeV$^{-2}$. A similar
compact formula may not exist for SDP model due to the non-exponential (power) terms in the relevant differential cross section (see, for instance Ref. \cite{Fiore:1998jx} for a discussion on the slope in SDP model). Concerning the NL model the slope of the forward cone is,
\begin{eqnarray}\label{slope}
B(s,Q^2,t)=\frac{d}{dt}\ln|A|^2=2\left[b_1+\ln\left({\frac{s}{s_0}}\right)\right]\frac{\alpha'}{1-\alpha_2
t}+ 2b_2\,\frac{\alpha'}{1-\alpha_2 z},
\end{eqnarray}
which, in the forward limit, $t=0$ reduces to
\begin{eqnarray}\label{slope1}
B(s,Q^2)=2\left[b_1+\ln\left(\frac{s}{s_0}\right)\right]\alpha'+
2b_2\,\frac{\alpha'}{1+\alpha_2 \widetilde{Q}^2},
\end{eqnarray}
where $s=W_{\gamma p}^2$. The slope $B_{NL}$ shows shrinkage in $s$ and antishrinkage in $\widetilde{Q}^2$. Numerically, one has $B_{\mathrm{NL}} \left(\langle W_{\gamma p}\rangle, \widetilde{Q}=m_{\rho} \right)  \approx 6.60 $ GeV$^{-2}$

The CMS Collaboration has parametrized the $t$-dependence using the simple Regge formula, $B(W_{\gamma p})=B_0+ 4\alpha^{\prime} \ln (W_{\gamma p}/W_0)$, with $W_0 = 92.6$ GeV, $\alpha^{\prime}=0.28 \pm 0.16$ GeV$^{-2}$ and $B_0\approx 10.3$ GeV$^{-2}$. data. On the other hand, H1 parametrized in the same way, with $W_0 = 40$ GeV, $\alpha^{\prime}=0.233\pm 0.064 (\mathrm{stat.})$ GeV$^{-2}$ and $B_0=9.59\pm 0.14\,(\mathrm{stat.})$ GeV$^{-2}$. Therefore, $B(92.6\,GeV) \approx 10.3$ GeV$^{-2}$ for CMS and 10.5 GeV$^{-2}$ for H1.

\begin{figure}[t]
\centering
 \includegraphics[width=.7\textwidth]{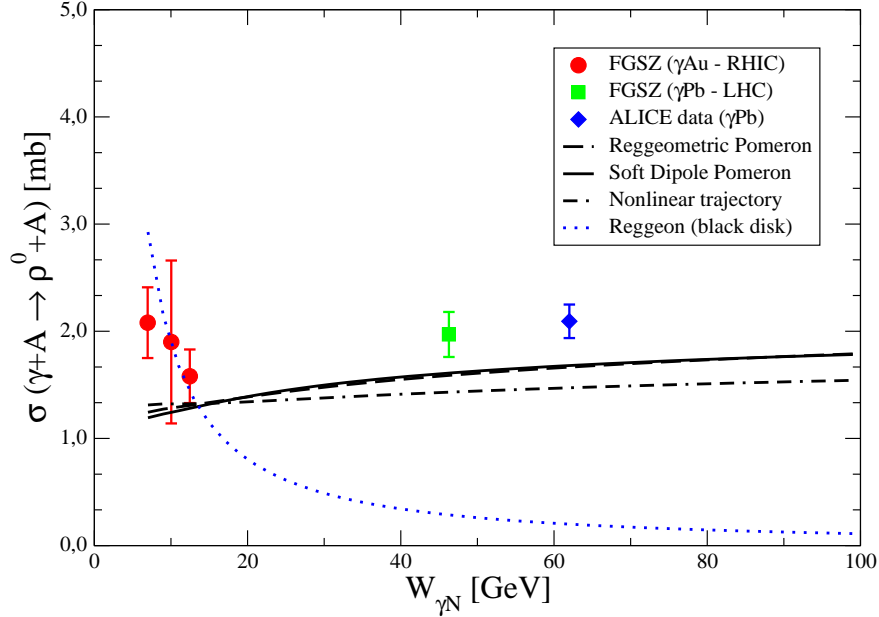}
 \caption{Nuclear coherent cross section as a function of centre-of-mass energy of the photon-nucleon system, $W_{\gamma N}$. Reggeometric Pomeron prediction is represented by dashed curve, the Soft Dipole by solid curve and that one for Nonlinear Trajectory by the dot-dashed curve. As an illustration, the dotted curve represents the low energy contribution. }
  \label{fig:photonuclear}
\end{figure}

Moving now to the $\rho^0$ production on nuclei, predictions to the coherent nuclear scattering cross section, $\sigma (\gamma +A \rightarrow \rho^0 + A)$, are presented in Fig. ~\ref{fig:photonuclear} as a function of photon-nucleon energy, $W_{\gamma N}$. A comparison is done with the extracted values of the coherent cross sections performed in Ref. \cite{Frankfurt:2015cwa} (labeled FGSZ) and by ALICE Collaboration in Ref. \cite{ALICE:2021jnv} using the measured data on UPCs at the LHC (PbPb collisions at 5.02 TeV). The low energy cross sections for $\gamma Au$ collisions (FGSZ) is also presented for completeness (RHIC AuAu collisions). The description is quite reasonable for the energy dependence and all models underestimate the extracted cross section by 30\%. The RP and SDP models produce almost identical results and the NL model gives a smaller normalization. It is clear that the similar results are due to the strong nuclear effects given that the calculated $\sigma (\gamma p\rightarrow \rho^0p)$ shown differences between models at large and low energies. As an illustration, at low energy we consider a black disk scaling following Ref. \cite{Klein:1999qj}, $\sigma(\gamma A \rightarrow  \rho^0 A)\approx A^{4/3}\sigma(\gamma p \rightarrow  \rho^0 p)=A^{4/3}\,(YW_{\gamma N}^{-\eta})$ (with $Y= 26$ $\mu$b and $\eta = 1.23$), which is represented by the dotted curve. In the calculations we added the reggeon contribution to the photoproduction off nucleons. The $\gamma Xe$ cross section has been extracted by ALICE Collaboration in Ref. \cite{ALICE:2021jnv} from the measured $d\sigma /dy = 131.50 \pm 24.96$ (errors summed into quadrature) in XeXe UPC collisions at 5.44 TeV. The cross section corresponds to the energy of photon-nucleon system $W_{\gamma N}=65$ GeV and $\sigma (\gamma Xe\rightarrow \rho^0Xe) = 1.12 \pm 0.21$ mb. The predicted values from the models are 1.07 mb (RP), 1.08 mb (SDP) and 0.92 mb (NL), respectively.

\begin{figure}[t]
\centering
 \includegraphics[width=.9\textwidth]{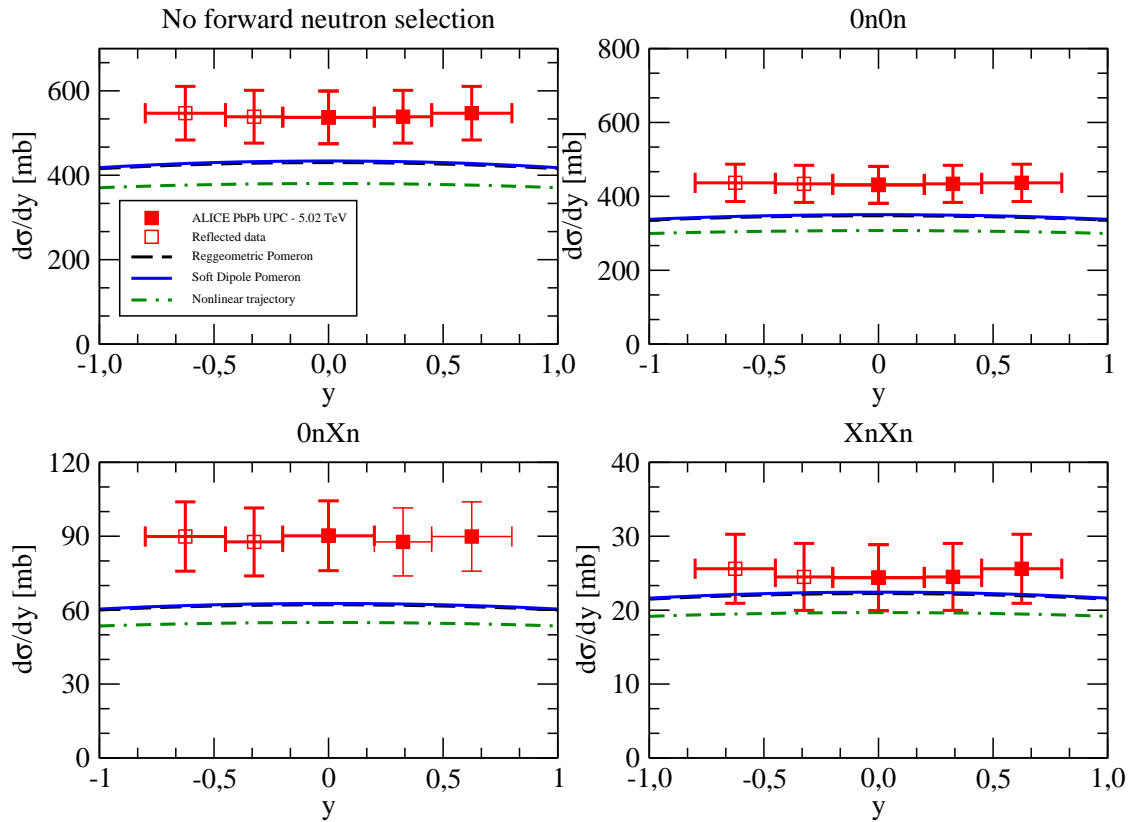}
 \caption{Rapidity distribution for the coherent photoproduction of $\rho^0$ mesons in PbPb UPC in midrapidity at $\sqrt{s_{\mathrm{NN}}} = 5.02$ TeV . Predictions for forward-neutron selection (top left), as well as for 0n0n (top right), 0nXn (bottom left)
and XnXn (bottom right) neutron class selections are presented. Data from ALICE Collaboration \cite{ALICE:2020ugp}.}
  \label{fig:ALICE502PbPb}
\end{figure}

The predictions for the nuclear coherent cross section presented in Fig. \ref{fig:photonuclear} and calculated by using Eq. (\ref{eq:sigtotgammaA}) will be used as input for computing the $\rho^0$ photoproduction is ultraperipheral PbPb and XeXe collisions at the LHC. The corresponding rapidity distribution in AA UPCs takes a factorized form in Equivalent Photon Approximation (EPA) and it is given by:
\begin{eqnarray}
\frac{d^2\sigma (A+A\rightarrow A+\rho^0+A)}{dy}=\omega_+\frac{dN_{\gamma/A}(\omega^+)}{d\omega}\sigma_{\gamma A\rightarrow \rho^0A}(\omega^+) +\omega^-\frac{dN_{\gamma/A}(\omega^-)}{d\omega}\sigma_{\gamma A\rightarrow \rho^0A}(\omega^-),
\end{eqnarray}
where $dN_{\gamma/A}(\omega)/d\omega$ is the photon flux in nucleus $A$ and $\omega$ is the photon energy. For fixed rapidity $y$ and transverse momentum $p_T^2 \approx |t|$ of the produced mesons, the photon momentum is given by $\omega^{\pm} = \frac{m_{\rho}^2-t}{2M_Te^{\mp}}$. $M_T$ is the transverse mass of the mesons. For simplicity, the analytical expression for the flux of photons produced by a fast-moving point-like charge has been considered \cite{Klein:1999qj,Baltz:2007kq}.

For large nuclei the process can be followed by additional photon exchanges between the colliding ions which lead to possible excitation of one or both of them. This can be accompanied by emission of neutrons moving along the direction of the ion beams \cite{Baltz:2002pp}. Usually, the additional photon exchanges can be accounted by an effective modification of the photon flux. Therefore, a selection can be done for the different channels, $k$, which are identified by emission of various number of neutrons. The neutron class
selections are for instance $k =$ 0n0n, 0nXn and XnXn. Thus, the photon flux for channel $k$ is expressed as \cite{Baltz:2002pp},
\begin{eqnarray}
\frac{dN_{\gamma/A}^k(\omega)}{d\omega} = \int_{2R_A}^{\infty}d^2\vec{b}\,\frac{d^3N_{\gamma/A}(\omega,\vec{b})}{d\omega d^2\vec{b}}P_k(\vec{b})
\end{eqnarray}
where $d^3N_{\gamma A}/d\omega d^2\vec{b}$ is the photon spectrum at a perpendicular distance $\vec{b}$ from the center of the emitting nucleus. The quantity $P_k$ is the probability to emit a given number of neutrons corresponding to the channel $k$. Here, we follow the modification of the photon flux due to the emission of the forward neutrons given in Ref. \cite{Baltz:2002pp}.
\begin{figure}[t]
\centering
 \includegraphics[width=.8\textwidth]{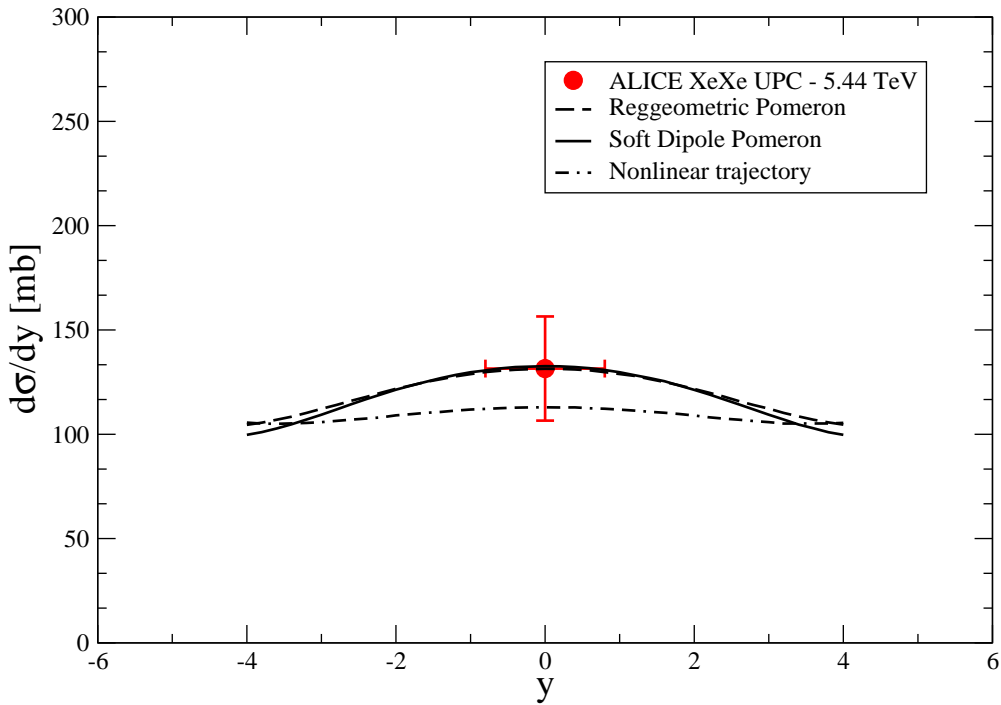}
 \caption{ Rapidity distribution for the coherent photoproduction of $\rho^0$ mesons in XeXe UPC in midrapidity at $\sqrt{s_{\mathrm{NN}}} = 5.44$ TeV. Predictions of the Reggeometric Pomeron model (dashed curve), Soft Dipole Pomeron model (solid curve) and nonlinear Pomeron trajectory (dot-dashed curve) are compared to data from ALICE Collaboration \cite{ALICE:2021jnv}.}
  \label{fig:ALICE544XeXe}
\end{figure}

In Fig. \ref{fig:ALICE502PbPb} results are shown for coherent nuclear $\rho^0$ production for PbPb peripheral collisions at $\sqrt{s_{\mathrm{NN}}} = 5.02$ TeV ($p_T<0.2$ GeV). The rapidity distribution of the meson is presented and compared to the ALICE collaboration measurements \cite{ALICE:2020ugp}. As the data refer to the absolute value of rapidity (filled squares) the data points at positive rapidities have been reflected into negative rapidities (open squares). In the top panel on the left the rapidity distribution is shown for no forward-neutron selection. The remaining plots present the cross section for coherent $\rho^0$ production accompanied by nuclear breakup. Thus, the following forward-neutron classes selection is shown: 0n0n, 0nXn and XnXn, respectively. The predictions for the RP model are labeled by the dashed curves, SDP model by the solid curves whereas the ones for the NL model are labeled by the dotted-dashed curves. The models underestimate the no forward neutron selection data within the error bars as expected based on results presented in Fig. \ref{fig:photonuclear} . The same occurs for the 0n0n and 0nXn classes. On the other hand, RP and SDP models describe XnXn cross section at midrapidity and NL result is close to measurements. Our calculations produce lower normalizations compared for instance to Guzey, Kryshen and Zhalov (GKZ) model \cite{Frankfurt:2015cwa}. It is based on a modified vector-dominance model where the interaction between the photon hadronic fluctuations and nucleons is described by the Gribov-Glauber formalism. Moreover, comparing our calculations to the STARLIGHT Monte Carlo \cite{Klein:2016yzr} the predictions are somewhat similar. This can be anticipated as the formalism for nuclear effect is the same in both cases. 

\begin{figure}[t]
\centering
 \includegraphics[width=.8\textwidth]{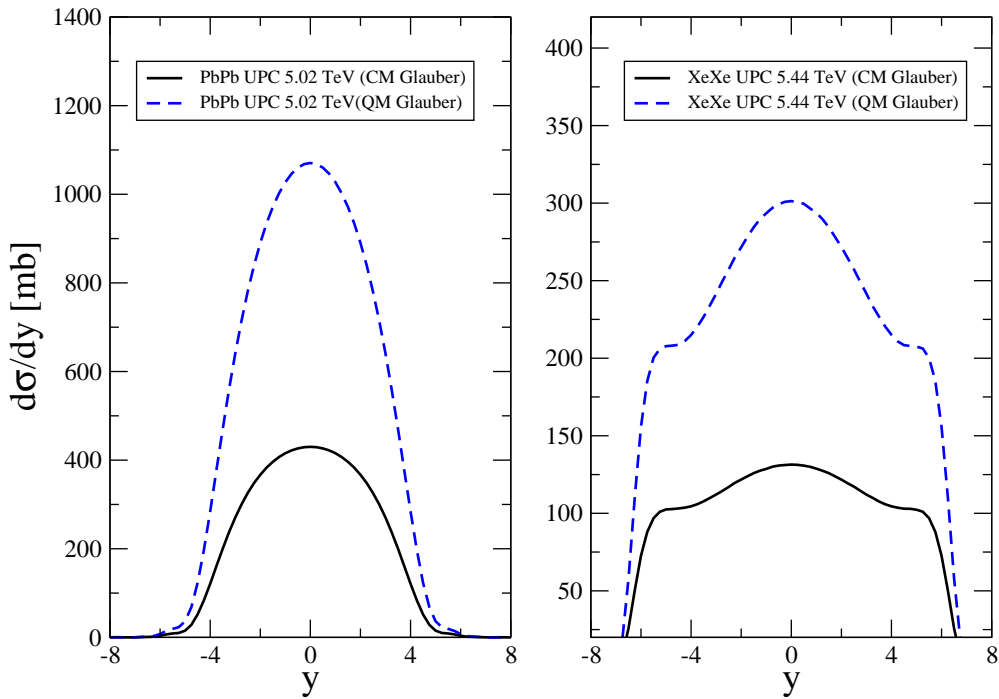}
 \caption{ Rapidity distribution for the coherent photoproduction of $\rho^0$ mesons in PbPb (5.02 TeV, left panel) and XeXe (5.44 TeV, right panel) UPCs at the LHC. Results for the Reggeometric Pomeron model only. Comparison of nuclear effects by using CM approach (solid curves) and GG approach (dashed curves) is done. }
  \label{fig:comp_Glauber}
\end{figure}

 In Fig. \ref{fig:ALICE544XeXe} results are shown for XeXe collisions at the energy of $\sqrt{s_{\mathrm{NN}}} = 5.44$ TeV. The measurement performed by ALICE Collaboration \cite{ALICE:2021jnv} at midrapidity is also presented. The notation for the curves is the same as previous figure. In general, the three models are suitable to predict the magnitude and shape of the rapidity distribution in XeXe UPCs. The nuclear effects for xenon nucleus are less intense as for lead nucleus and the cross section normalization now is compatible with data. We have shown before that the coherent cross section $\sigma (\gamma Xe\rightarrow \rho^0Xe)$ is correctly described by the models discussed here.

Finally, some comments on the nuclear effects are in order. As presented  in section 2, our calculations follow the STARLIGHT MC  generator approach. There the total cross section of the $\rho^0A$ interaction is obtained from classical mechanics formula (CM) instead of the quantum mechanics (GG) one. It was estimated that  $\frac{\sigma_{tot}^{\mathrm{GG}}(\rho^0Pb)}{\sigma_{tot}^{\mathrm{CM}}(\rho^0Pb) }\approx 1.7$ and  $\frac{\sigma_{tot}^{\mathrm{GG}}(\rho^0Xe)}{\sigma_{tot}^{\mathrm{CM}}(\rho^0Xe) } \approx 1.5$. Thus, the deviation in midrapidity at the LHC energies is very large as one can see in Fig. \ref{fig:comp_Glauber}. In the figure, predictions are made by using CM (solid curves) and GG (dashed curves)) approaches and Reggeometric Pomeron model. Left plot corresponds to PbPb UPCs and right plot to the XeXe UPCs at LHC energies.

In Fig. \ref{fig:comp_Glauber} the rapidity range is extended to forward/backward rapidities and the calculations include reggeons. In the rapidity range (midrapidity) considered in this work the Pomeron exchange is the main contribution and reggeon exchange piece is suppressed. From Fig. \ref{fig:photonuclear} it is clear that the reggeon contribution turns to be relevant for energies $W_{\gamma N}\lesssim 10 $ GeV. This can be translated into the corresponding rapidity.  The rapidity of the $\rho^0$ meson is related to the center-of-mass energy of the photon-nucleus system through, $W_{\gamma N}^2 = \sqrt{s_{\mathrm{NN}}}M_{\rho}e^{y}$. A simple calculation gives $|y|\simeq 3.66$ for lead and $|y|\simeq 3.74$ for xenon. The structure of the rapidity distributions represents an interplay of several phenomena. Namely, the energy dependence of the photon-nucleon cross section, suppression due to nuclear shadowing, and descend of the flux of high-energy photons drive the distribution in the central and forward (backward) rapidity regions. The bumps or shoulders at large $|y|$ are due to an enhanced contribution of low-energy
photoproduction related to the secondary reggeon exchange in the meson-nucleon interaction.The nuclear shadowing at low energies is more intense for lead ($A=208$) nuclei compared to the xenon ($A=129$) ones. This is the reason for the shoulder appearing in XeXe and not in PbPb collisions in the region $|y|\simeq 4$.

\section{Conclusions}

In this work predictions for exclusive $\rho^0$ photoproduction in $ep$ collisions and also UPCs collisions at the LHC are presented and compared with the recent experimental measurements. The theoretical approach is based on Regge phenomenology and 3 different models have been considered: i) the single-component Reggeometric Pomeron model, ii) the Soft Dipole Pomeron model and iii) a factorized Regge model with nonlinear Pomeron trajectory. The comparison of the results to the H1 data has shown that the models are suitable for the small-$|t|$ region and clear deviations at large $t$. It is worth mentioning that the models parameters have been tuned to describe previous measurements by DESY-HERA. Updated fits including new data sets would be helpful. Concerning the rapidity distributions in PbPb UPCs, RP and SDP models are in agreement with data in case of no forward neutron selection whereas NL model underestimates the cross section. In the case of XeXe UPCs all models are consistent with the data following the same trend observed in PbPb collisions. 

The present paper generalizes results on $\rho$ photoproduction in two
directions. First, we have collected, compared, critically revised and partly
refitting three different Regge-pole models of exclusive $\rho^0$ photoproduction. An important conclusion from this
comparison is the the importance of the
non-linear nature of Regge trajectories, including that of the Pomeron.
This is seen explicitly e.g. in Fig. \ref{fig:ALICE544XeXe}.
Non-linearity of Regge trajectories is seen both at small $|t|$, manifest
as a "break" in the differential cross section,
of elastic proton-proton scattering as well as at moderate and large
values of $|t|$, matching transition to hard scattering,
described by perturbative QCD. Construction of explicit models of
non-linear Regge trajectories compatible with analytic and
asymptotic constrains of the theory is an important and permanently
developing field of research. We intend to continue research in
this direction by including also DIS, DVCS and production of other vector
mesons.

Secondly, we continue investigation of how models of vector meson
production in $ep$ scattering affect the results in
ultra-peripheral nuclear collisions. This direction of research is
especially promising also because of the planned experiments
at future accelerators.

 \section*{Acknowledgements}
We are grateful to Vladyslav Libov for collaborating at an earlier stage of the work. L.J. was supported by the NASU grant 1230/22-1 \textit{Fundamental Properties of Matter}.
MVTM was supported by funding agencies CAPES (Finance Code 001) and CNPq (grant number 306101/2018-1), Brazil.


\begin{thebibliography}{99}
\bibitem{H1:2020lzc}
V.~Andreev \textit{et al.} [H1],
Eur. Phys. J. C \textbf{80}, no.12, 1189 (2020)
doi:10.1140/epjc/s10052-020-08587-3
[arXiv:2005.14471 [hep-ex]].

\bibitem{CMS:2019awk}
A.~M.~Sirunyan \textit{et al.} [CMS],
Eur. Phys. J. C \textbf{79}, no.8, 702 (2019)
doi:10.1140/epjc/s10052-019-7202-9
[arXiv:1902.01339 [hep-ex]].

\bibitem{ALICE:2020ugp}
S.~Acharya \textit{et al.} [ALICE],
JHEP \textbf{06}, 035 (2020)
doi:10.1007/JHEP06(2020)035
[arXiv:2002.10897 [nucl-ex]].

\bibitem{ALICE:2021jnv}
S.~Acharya \textit{et al.} [ALICE],
Phys. Lett. B \textbf{820}, 136481 (2021)
doi:10.1016/j.physletb.2021.136481
[arXiv:2101.02581 [nucl-ex]].

\bibitem{hdqcd} 
  F.~Gelis, E.~Iancu, J.~Jalilian-Marian and R.~Venugopalan,
    Ann.\ Rev.\ Nucl.\ Part.\ Sci.\  {\bf 60}, 463 (2010);
  H.~Weigert,  Prog.\ Part.\ Nucl.\ Phys.\  {\bf 55}, 461 (2005); J.~Jalilian-Marian and Y.~V.~Kovchegov, Prog.\ Part.\ Nucl.\ Phys.\  {\bf 56}, 104 (2006).

\bibitem{Morreale:2021pnn}
A.~Morreale and F.~Salazar,
Universe \textbf{7}, no.8, 312 (2021)
doi:10.3390/universe7080312
[arXiv:2108.08254 [hep-ph]].

\bibitem{nik} N. N. Nikolaev, B. G. Zakharov,  Phys. Lett. B  {\bf 332}, 184 (1994); {Z. Phys. C} {\bf 64}, 631 (1994).

\bibitem{Nemchik:1996pp} 
  J.~Nemchik, N.~N.~Nikolaev, E.~Predazzi and B.~G.~Zakharov,
  Phys.\ Lett.\ B {\bf 374}, 199 (1996).

\bibitem{Klein:2020nvu}
S.~Klein, D.~Tapia Takaki, J.~Adam, C.~Aidala, A.~Angerami, B.~Audurier, C.~Bertulani, C.~Bierlich, B.~Blok and J.~D.~Brandenburg, \textit{et al.}
[arXiv:2009.03838 [hep-ph]].


\bibitem{Goncalves:2020cir}
V.~P.~Gon\c{c}alves and B.~D.~Moreira,
Eur. Phys. J. C \textbf{80}, no.6, 492 (2020)
doi:10.1140/epjc/s10052-020-8043-2
[arXiv:2003.11438 [hep-ph]].

\bibitem{Forshaw:1999uf}
J.~R.~Forshaw, G.~Kerley and G.~Shaw,
Phys. Rev. D \textbf{60}, 074012 (1999)
doi:10.1103/PhysRevD.60.074012
[arXiv:hep-ph/9903341 [hep-ph]].


\bibitem{Fazio:2013uwa}
S.~Fazio, R.~Fiore, A.~Lavorini, L.~Jenkovszky and A.~Salii,
Acta Phys. Polon. B \textbf{44}, 1333-1353 (2013)
doi:10.5506/APhysPolB.44.1333
[arXiv:1304.1891 [hep-ph]].

\bibitem{Fazio:2013hza}
S.~Fazio, R.~Fiore, L.~Jenkovszky and A.~Salii,
Phys. Rev. D \textbf{90}, no.1, 016007 (2014)
doi:10.1103/PhysRevD.90.016007
[arXiv:1312.5683 [hep-ph]].


\bibitem{Klein:2020fmr}
S.~Klein and P.~Steinberg,
Ann. Rev. Nucl. Part. Sci. \textbf{70}, 323-354 (2020)
doi:10.1146/annurev-nucl-030320-033923
[arXiv:2005.01872 [nucl-ex]].



\bibitem{Jenkovszky:2018itd}
L.~Jenkovszky, R.~Schicker and I.~Szanyi,
Int. J. Mod. Phys. E \textbf{27}, no.08, 1830005 (2018)
doi:10.1142/S0218301318300059
[arXiv:1902.05614 [hep-ph]].



\bibitem{Jenkovszky:1996hk}
L.~L.~Jenkovszky, E.~S.~Martynov and F.~Paccanoni,
[arXiv:hep-ph/9608384 [hep-ph]].


\bibitem{Fiore:1998jx}
R.~Fiore, L.~L.~Jenkovszky and F.~Paccanoni,
Eur. Phys. J. C \textbf{10}, 461-467 (1999)
doi:10.1007/s100520050768
[arXiv:hep-ph/9812458 [hep-ph]].

\bibitem{Martynov:2002ez}
E.~Martynov, E.~Predazzi and A.~Prokudin,
Phys. Rev. D \textbf{67}, 074023 (2003)
doi:10.1103/PhysRevD.67.074023
[arXiv:hep-ph/0207272 [hep-ph]].

\bibitem{Fazio:2011ex}
S.~Fazio, R.~Fiore, L.~Jenkovszky and A.~Lavorini,
Phys. Rev. D \textbf{85}, 054009 (2012)
doi:10.1103/PhysRevD.85.054009
[arXiv:1109.6374 [hep-ph]].

\bibitem{Fiore:2014oha}
R.~Fiore, L.~Jenkovszky, V.~Libov and M.~Machado,
Teor. Mat. Fiz. \textbf{182}, no.1, 171-181 (2014)
doi:10.1007/s11232-015-0252-8
[arXiv:1408.0530 [hep-ph]].

\bibitem{Fiore:2014lxa}
R.~Fiore, L.~Jenkovszky, V.~Libov, M.~V.~T.~Machado and A.~Salii,
[arXiv:1506.01990 [hep-ph]]. Contribution to: Diffraction 2014.

\bibitem{Fiore:2015yya}
R.~Fiore, L.~Jenkovszky, V.~Libov, M.~V.~T.~Machado and A.~Salii,
AIP Conf. Proc. \textbf{1654}, no.1, 090002 (2015)
doi:10.1063/1.4916009

\bibitem{Jenkovszky:2021sis}
L.~Jenkovszky, V.~Libov and M.~V.~T.~Machado,
Phys. Lett. B \textbf{824}, 136836 (2022).
doi:10.1016/j.physletb.2021.136836
[arXiv:2111.13389 [hep-ph]].

\bibitem{Jenkovszky:2022wcw}
L.~Jenkovszky, V.~Libov and M.~V.~T.~Machado,
Phys. Lett. B \textbf{827}, 137004 (2022)
doi:10.1016/j.physletb.2022.137004
[arXiv:2202.02162 [hep-ph]].

\bibitem{Accardi:2012qut}
A.~Accardi, J.~L.~Albacete, M.~Anselmino, N.~Armesto, E.~C.~Aschenauer, A.~Bacchetta, D.~Boer, W.~K.~Brooks, T.~Burton and N.~B.~Chang, \textit{et al.}
Eur. Phys. J. A \textbf{52}, no.9, 268 (2016)
doi:10.1140/epja/i2016-16268-9
[arXiv:1212.1701 [nucl-ex]].

\bibitem{Anderle:2021wcy}
D.~P.~Anderle, V.~Bertone, X.~Cao, L.~Chang, N.~Chang, G.~Chen, X.~Chen, Z.~Chen, Z.~Cui and L.~Dai, \textit{et al.}
Front. Phys. (Beijing) \textbf{16}, no.6, 64701 (2021)
doi:10.1007/s11467-021-1062-0
[arXiv:2102.09222 [nucl-ex]].

\bibitem{Andre:2022xeh}
K.~D.~J.~Andr\'e, L.~Aperio Bella, N.~Armesto, S.~A.~Bogacz, D.~Britzger, O.~S.~Br\"uning, M.~D'Onofrio, E.~G.~Ferreiro, O.~Fischer and C.~Gwenlan, \textit{et al.}
Eur. Phys. J. C \textbf{82}, no.1, 40 (2022)
doi:10.1140/epjc/s10052-021-09967-z
[arXiv:2201.02436 [hep-ex]].

\bibitem{Henkels:2022bne}
C.~Henkels, E.~G.~de Oliveira, R.~Pasechnik and H.~Trebien,
[arXiv:2207.13756 [hep-ph]].

\bibitem{Guzey:2020pkq}
V.~Guzey, E.~Kryshen and M.~Zhalov,
Phys. Rev. C \textbf{102}, no.1, 015208 (2020)
doi:10.1103/PhysRevC.102.015208
[arXiv:2002.09683 [hep-ph]].

\bibitem{Ma:2019mwr}
Z.~L.~Ma, Z.~Lu, J.~Q.~Zhu and L.~Zhang,
Phys. Rev. D \textbf{104}, no.7, 074023 (2021)
doi:10.1103/PhysRevD.104.074023
[arXiv:1910.04509 [hep-ph]].



\bibitem{Guzey:2018bay}
V.~Guzey, E.~Kryshen and M.~Zhalov,
Phys. Lett. B \textbf{782}, 251-255 (2018)
doi:10.1016/j.physletb.2018.05.058
[arXiv:1803.07638 [hep-ph]].

\bibitem{Goncalves:2018blz}
V.~P.~Gon\c{c}alves, F.~S.~Navarra and D.~Spiering,
Phys. Lett. B \textbf{791}, 299-304 (2019)
doi:10.1016/j.physletb.2019.03.007
[arXiv:1811.09124 [hep-ph]].

\bibitem{Xie:2022sjm}
Y.~P.~Xie and V.~P.~Goncalves,
Phys. Rev. D \textbf{105}, no.1, 014033 (2022)
doi:10.1103/PhysRevD.105.014033
[arXiv:2201.10499 [hep-ph]].


\bibitem{Xing:2020hwh}
H.~Xing, C.~Zhang, J.~Zhou and Y.~J.~Zhou,
JHEP \textbf{10}, 064 (2020)
doi:10.1007/JHEP10(2020)064
[arXiv:2006.06206 [hep-ph]].

\bibitem{Demirci:2022wuy}
S.~Demirci, T.~Lappi and S.~Schlichting,
[arXiv:2206.05207 [hep-ph]].


\bibitem{Cisek:2022yjj}
A.~Cisek, W.~Sch\"afer and A.~Szczurek,
[arXiv:2209.06578 [hep-ph]].

\bibitem{Mantysaari:2022bsp}
H.~M\"antysaari and J.~Penttala,
Phys. Rev. D \textbf{105}, no.11, 11 (2022)
doi:10.1103/PhysRevD.105.114038
[arXiv:2203.16911 [hep-ph]].

\bibitem{Cepila:2018zky}
J.~Cepila, J.~G.~Contreras, M.~Krelina and J.~D.~Tapia Takaki,
Nucl. Phys. B \textbf{934}, 330-340 (2018)
doi:10.1016/j.nuclphysb.2018.07.010
[arXiv:1804.05508 [hep-ph]].

\bibitem{Bendova:2018bbb}
D.~Bendova, J.~Cepila and J.~G.~Contreras,
Phys. Rev. D \textbf{99}, no.3, 034025 (2019)
doi:10.1103/PhysRevD.99.034025
[arXiv:1811.06479 [hep-ph]].



\bibitem{Khoze:2019xke}
V.~A.~Khoze, A.~D.~Martin and M.~G.~Ryskin,
J. Phys. G \textbf{46}, no.8, 085002 (2019)
doi:10.1088/1361-6471/ab2009
[arXiv:1902.08136 [hep-ph]].


\bibitem{H1:1996prv}
S.~Aid \textit{et al.} [H1],
Nucl. Phys. B \textbf{463}, 3-32 (1996)
doi:10.1016/0550-3213(96)00045-4
[arXiv:hep-ex/9601004 [hep-ex]].

\bibitem{ZEUS:1997rof}
J.~Breitweg \textit{et al.} [ZEUS],
Eur. Phys. J. C \textbf{2}, 247-267 (1998)
doi:10.1007/s100520050136
[arXiv:hep-ex/9712020 [hep-ex]].

\bibitem{ZEUS:1995bfs}
M.~Derrick \textit{et al.} [ZEUS],
Z. Phys. C \textbf{69}, 39-54 (1995)
doi:10.1007/s002880050004
[arXiv:hep-ex/9507011 [hep-ex]].



\bibitem{Klein:1999qj}
S.~Klein and J.~Nystrand,
Phys. Rev. C \textbf{60}, 014903 (1999)
doi:10.1103/PhysRevC.60.014903
[arXiv:hep-ph/9902259 [hep-ph]].

\bibitem{Nystrand:1998hw}
J.~Nystrand \textit{et al.} [STAR],
[arXiv:nucl-ex/9811007 [nucl-ex]].

\bibitem{Klein:2016yzr}
S.~R.~Klein, J.~Nystrand, J.~Seger, Y.~Gorbunov and J.~Butterworth,
Comput. Phys. Commun. \textbf{212}, 258-268 (2017)
doi:10.1016/j.cpc.2016.10.016
[arXiv:1607.03838 [hep-ph]].

\bibitem{Bauer:1977iq}
T.~H.~Bauer, R.~D.~Spital, D.~R.~Yennie and F.~M.~Pipkin,
Rev. Mod. Phys. \textbf{50}, 261 (1978)
[erratum: Rev. Mod. Phys. \textbf{51}, 407 (1979)]
doi:10.1103/RevModPhys.50.261



\bibitem{Frankfurt:2003wv}
L.~Frankfurt, M.~Strikman and M.~Zhalov,
Acta Phys. Polon. B \textbf{34}, 3215-3254 (2003)
[arXiv:hep-ph/0304301 [hep-ph]].



\bibitem{Davies:1976zzb}
K.~T.~R.~Davies and J.~R.~Nix,
Phys. Rev. C \textbf{14}, 1977-1994 (1976)
doi:10.1103/PhysRevC.14.1977

\bibitem{Frankfurt:2002sv}
L.~Frankfurt, M.~Strikman and M.~Zhalov,
Phys. Rev. C \textbf{67}, 034901 (2003)
doi:10.1103/PhysRevC.67.034901
[arXiv:hep-ph/0210303 [hep-ph]].

\bibitem{Frankfurt:2002wc}
L.~Frankfurt, M.~Strikman and M.~Zhalov,
Phys. Lett. B \textbf{537}, 51-61 (2002)
doi:10.1016/S0370-2693(02)01882-8
[arXiv:hep-ph/0204175 [hep-ph]].

\bibitem{Guzey:2013xba}
V.~Guzey, E.~Kryshen, M.~Strikman and M.~Zhalov,
Phys. Lett. B \textbf{726}, 290-295 (2013)
doi:10.1016/j.physletb.2013.08.043
[arXiv:1305.1724 [hep-ph]].



\bibitem{Frankfurt:2015cwa}
L.~Frankfurt, V.~Guzey, M.~Strikman and M.~Zhalov,
Phys. Lett. B \textbf{752}, 51-58 (2016)
doi:10.1016/j.physletb.2015.11.012
[arXiv:1506.07150 [hep-ph]].




\bibitem{Baltz:2007kq}
A.~J.~Baltz, G.~Baur, D.~d'Enterria, L.~Frankfurt, F.~Gelis, V.~Guzey, K.~Hencken, Y.~Kharlov, M.~Klasen and S.~R.~Klein, \textit{et al.}
Phys. Rept. \textbf{458}, 1-171 (2008)
doi:10.1016/j.physrep.2007.12.001
[arXiv:0706.3356 [nucl-ex]].




\bibitem{Baltz:2002pp}
A.~J.~Baltz, S.~R.~Klein and J.~Nystrand,
Phys. Rev. Lett. \textbf{89}, 012301 (2002)
doi:10.1103/PhysRevLett.89.012301
[arXiv:nucl-th/0205031 [nucl-th]].




\end{thebibliography}
\end{document}